# Asymptotic Multi-Layer Analysis of Wind Over Unsteady Monochromatic Surface Waves *

S.G. Sajjadi[1,3], J.C.R. Hunt[2,3] and F. Drullion[1]
*(1) Department of Mathematics, Embry-Riddle Aeronautical University, FL, USA.*
*(2) Earth Sciences, University College London, UK.*
*(3) Trinity College, University of Cambridge, UK.*

**Abstract.** Asymptotic multi-layer analyses and computation of solutions for turbulent flows over steady and unsteady monochromatic surface wave are reviewed, in the limits of low turbulent stresses and small wave amplitude. The structure of the flow is defined in terms of asymptotically-matched thin-layers, namely the surface layer and a critical layer, whether it is 'elevated' or 'immersed', corresponding to its location above or within the surface layer. The results particularly demonstrate the physical importance of the singular flow features and physical implications of the elevated critical layer in the limit of the unsteadiness tending to zero. These agree with the variational mathematical solution of Miles [1] for small but finite growth rate, but they are not consistent physically or mathematically with his analysis in the limit of growth rate tending to zero. As this and other studies conclude, in the limit of zero growth rate the effect of the elevated critical layer is eliminated by finite turbulent diffusivity, so that the perturbed flow and the drag force are determined by the asymmetric or sheltering flow in the surface shear layer and its matched interaction with the upper region. But for groups of waves, in which the individual waves grow and decay, there is a net contribution of the elevated critical layer to the wave growth. Critical layers, whether elevated or immersed, affect this asymmetric sheltering mechanism, but in quite a different way to their effect on growing waves. These asymptotic multi-layer methods lead to physical insight and suggest approximate methods for analysing higher amplitude and more complex flows, such as flow over wave groups.

## 1. Introduction

Various mechanisms have been proposed and evaluated to describe turbulent winds over water waves and thence to explain how such waves are generated. But despite 100 years of theoretical research and more recently, detailed measurements and numerical computations, the nature of these mechanisms and their relative magnitude remain controversial even for the ideal case of monochromatic waves. Conferences of wave experts concluded [2,3] that more research is necessary even on ideal cases of steady and unsteady waves in order to resolve these controversies, and to improve in the forecasting of waves and their effects on weather, climate and forces on ocean structures.

---

\* For special issue of *J. Eng Math.* to honour Milton Van Dyke, edited by Len Schwartz.

In this paper, written in honour of the late Milton Van Dyke, it is shown that the different mechanisms affecting air flow over waves and their relative contributions can be quantified and understood by using the modern methods of asymptotic multi-layer (AML) analysis [4.5]. An advantage of AML methods is that they also indicate how different mechanisms can affect each other through weak non-linear interactions.

Many authors have applied AML methods to turbulent flows, including those over complex surfaces; first by making statistical assumptions, for example by deriving approximate equations for statistical moments, (usually first (i.e. mean) and second-order moments), and then assuming the asymptotic limit of $\varepsilon \to 0$, where $\varepsilon$ is the ratio, $U_*/U_0$. Here $U_*$ is the square root of the Reynolds shear stress or 'friction velocity', and $U_0$ is the undisturbed mean velocity. Another asymptotic assumption for the analysis of waves is that their slope $a/L$ is also very small, i.e. $a/L \to 0$, where $a$ is the wave height and $L$ is its wavelength. The third asymptotic limit used in calculating wind over waves is related to the unsteadiness of the mean flow and the wave surface, where the perturbations grow in proportion to $\exp(c_i t/L)$, where $c_i/U_0 > 0$ and $c_i$ is the complex part of the wave speed [6]. The equations used in these analyses depend on the turbulence modelling used, as this paper demonstrates.

The analysis of wind over waves – which is usually expressed in a coordinate frame moving with the wave speed $c_r$ also requires considering its multi-layer structure-defined in fig. 1 – which varies depending on the key variable of the wave speed relative to the friction velocity $c_r/U_*$. The asymptotic layers are located at the following levels:

(i) critical layer above the waves surface $z_c$, where $U(z_c) = c_r$, and $U(z)$ is the undisturbed velocity profile;

(ii) surface shear layer extending upwards from the surface ($z = 0$) over a thickness $l_s$. Note that at the bottom of this layer there may be an inner layer with thickness $l_i$. In flows where $c_r \ll U_*$ the critical layer at $z_c$ may be 'immersed' within this layer.

The first models of how air flow leads to wave growth were based on the intrinsic instability of air flow over water surfaces. For a steady flow this is the well known Kelvin-Helmholtz (KH) instability; the flow may be gusty and unsteady leading to unsteady KH waves [7]. Another kind of instability mechanism is driven by growing fluctuations in the air caused by instability in the boundary-layer air flow.

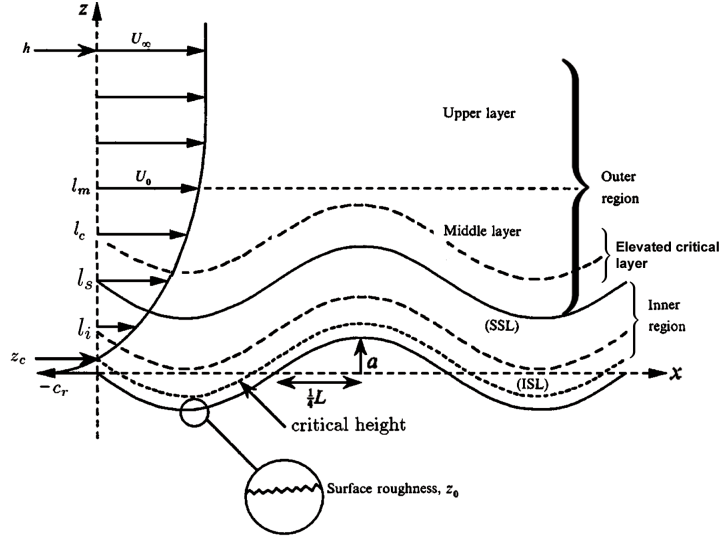

*Figure 1.* Schematic diagram for flow geometry and asymptotic multi-layer structure for analysing turbulent shear flow over steady and unsteady monochromatic waves.

These mechanisms are observed when wind is initiated over a flat water surface [8]. It is then observed that waves propagate with a small normalised rate of growth i.e. $c_i/U_* \ll 1$. Note that as waves grow they have a wide distribution of wavelengths $L$, travelling at different speeds. Generally they tend to form wave-groups.

Two main types of ideal model have been developed for slowly changing waves, which we compare in detail in this paper. Models for separated flow over high slopes ($a/L \sim 1$) [9] and, for low slopes ($a/L \ll 1$) [10,11,12] where $c_i = 0$ have been based on concepts similar to flows over hills, with the wind profile decelerating more on the downwind side than on the upwind side and an acceleration of mean flow over the top [6]. AML models enable the different processes to be calculated and compared [13]. The dominant mechanism for low slope waves is referred to as non-separated sheltering (NSS), which showed how the mean velocity shear could lead to significant energy transfer even when $a/L \ll 1$. Note that a critical layer exists in these flows since $c_r > 0$, as is observed in experiments and simulations [14,15], but its effect is small because (as shown in sections 2 and 3) the simulations near the critical layer over non-growing waves where $0 < c_i <\sim U_*$ do not correspond to the analytical or computed results of wind over growing waves for $0 < c_i \ll U_*$.

Because the steady state model of Jeffreys [9], which effectively assumed $a/L \sim 1$, could not explain the transfer of energy into typical waves with low slope, a new theory was developed by Miles [1] by considering how the mean velocity profile $U(z)$ over waves with low slope ($a/L \ll 1$) could lead to a significant transfer of energy provided the waves are growing slowly, i.e. $U_* \gg c_i > 0$. Miles' inviscid analysis, which centred on the critical layer, was not based on consistent AML methods, but rather a selective mixture of methods. As argued here in section 2 (and by Mastenbroek [11]), Miles' analysis led to a dubious conclusion (which has never been tested by rigorous numerical simulation) that, when $c_i \to 0$, the unsteady result can be applied to waves where $c_i = 0$.

Nevertheless, with some empirical adjustment, this critical layer (CL) model (which was 'justified' using approximate unsteady vortex dynamics by Lighthill [16]) is used by oceanographers and meteorologists around the world even for quite complex wave fields, which are far from the idealised form of the theory [17,18].

A number of authors have discussed combining NSS and unsteady CL mechanisms, perhaps applied to realistic groups of waves [19,13,20]. This will be considered in more detail in a later paper, using the AML methods described here.

## 2. Steady and unsteady linear analysis of outer region flow above the surface layer

We consider here the perturbation $\Delta u$ to the mean shear flow $U(z)$ in the outer region over unsteady monochromatic two-dimensional waves above the surface layer, i.e. $z > l_s$, where $l_s/L \ll 1$, in the limit in which $\varepsilon \ll 1$.

The wave surface $z_s$ has wavelength $L = 2\pi/k$ moving with speed $c_r$, so that
$$z_s = a \exp\{ik(x - c_r t) + kc_i t\}.$$

The height of the critical layer $z_c$, where $U(z_c) = c_r$, may or may not be above $l_s$.

The two-dimensional mean velocity field is defined by $\boldsymbol{u} = (U + \Delta u, \Delta w)$, where by continuity
$$\frac{\partial \Delta u}{\partial x} + \frac{\partial \Delta w}{\partial z} = 0.$$

The perturbation velocity $\Delta w = \mathscr{W}(z)e^{ik(x - c_r t) + kc_i t}$ is determined by the linearised momentum equation, in a frame of reference moving with the wave, where the amplitude of the perturbation, $\mathscr{W}$ satisfies the inhomogeneous Rayleigh equation

$$\frac{\partial^2 \mathscr{W}}{\partial z^2} - \left(k^2 + \frac{U''}{\mathscr{U} - ic_i}\right)\mathscr{W} = -\frac{i}{k(\mathscr{U} - ic_i)}\frac{\partial^2}{\partial z^2}\left(\nu_e \frac{\partial^2 \mathscr{W}}{\partial z^2}\right) \quad (1)$$

where $\nu_e$ is the eddy viscosity in the critical layer, and $\mathscr{U}(z) \equiv U(z) - c_r$. Note that if $\nu_e = 0$, the equation (1) is the Rayleigh equation and is singular at the critical height $z_c$ if the waves do not grow i.e. when $c_i = 0$. We note that when $\nu_e \neq 0$ this is a truncated form of the Orr-Sommerfeld equation. We remark that weak turbulence in the outer region produces second order stresses, whose effects on the waves are considered in section 3 (see also [13]).

For non-zero value of $\nu_e$, the solution to equation (1) is determined by the boundary condition at $z \sim l_s$, which is defined by matching between the outer region and the surface shear layer.

As with many unsteady shear flow problems, the basic mechanisms are best explained by considering the singularities of the governing equations, which is associated with the term $U''/\{U(z) - (c_r + ic_i)\}$ as $c_i \to 0$. For a typical monotonic mean wind profile with $U(z) = (U_*/\kappa)\ln(z/z_0)$ and vorticity $\omega = U_*/(z\kappa)$, where $\kappa$ is von Kármán constant, the peak velocity gradient is at the wave surface. However, the most significant perturbation to the mean vorticity occurs where the wave displacement leads to closed streamlines near $z = z_c$ [6], see fig. 2.

The local analysis near $z_c$, shows how the vertical profile of the in-phase and out of phase perturbation velocity, and pressure perturbation has a singular behavior near $z = z_c$ when $0 < c_i \ll 1$. Belcher et al. [21] showed that the local solution in the critical layer is[1]

$$\mathscr{W} \sim \{\mathscr{U}(z) - ic_i\}\left\{A + B\int^{z-z_c}\frac{d\zeta}{[\mathscr{U}(\zeta) - ic_i]^2}\right\} \quad (2)$$

where $A$ and $B$ are constants and can be determined by matching the inner and the outer solutions. They showed that in the limit of slow-growing waves $\varepsilon = c_i U_c''/U_c'^2 \ll 1$ (where the suffix $c$ refers to evaluation at the critical point where $U = c_r$) and a logarithmic profile for the mean velocity $U(z)$ the integral ($I$, say) in (2) can be reduced to

$$I \sim \frac{\varepsilon}{c_i U'_c} \ln(\xi - i) \quad \text{as } \xi \to \pm\infty$$
$$= \frac{\varepsilon}{2c_i U'_c}[\ln(\xi^2 + 1) + 2i\theta] \tag{3}$$

where $\xi = \zeta U'_c/c_i$, $\zeta = z - z_c$ and $\theta$ is given by

$$\tan\theta = -\xi^{-1} = -c_i/U'_c(z - z_c) \tag{4}$$

---

[1] This assumes that in the middle layer the advection term is negligible compared with the curvature term and thus (1) reduces to $\mathscr{W}'' - U''/(\mathscr{U} - ic_i)\mathscr{W} \sim 0$.

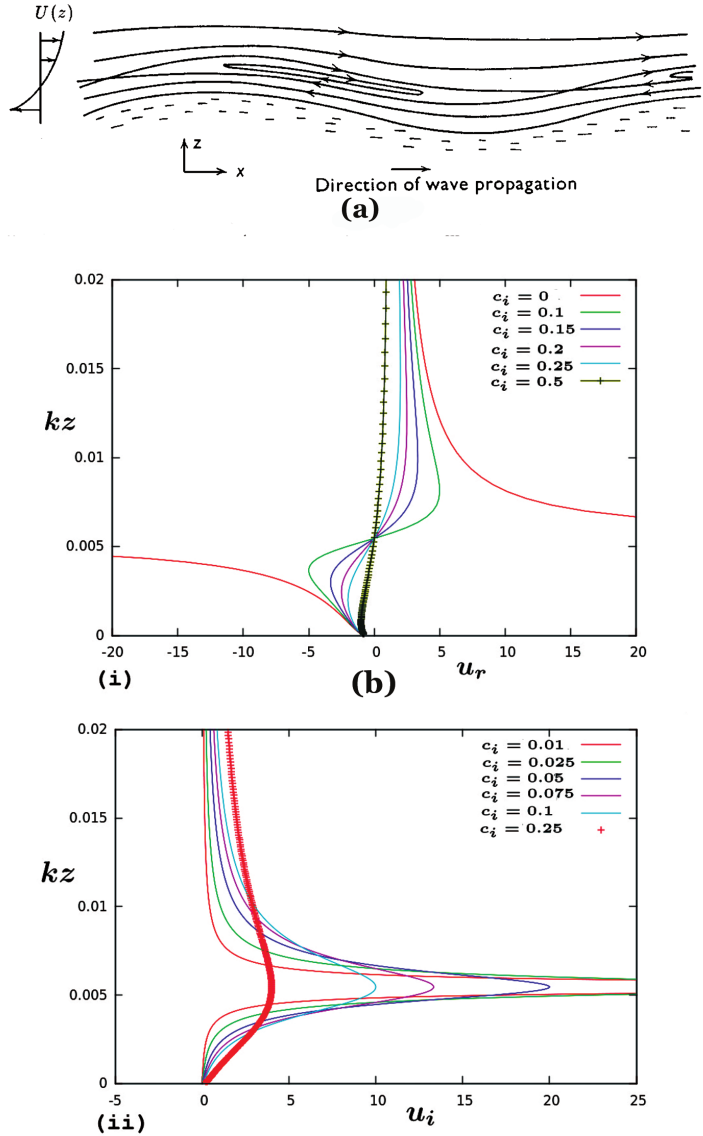

*Figure 2.* (a) Mean streamlines of flow over waves viewed as moving with the waves, whose closed loops are centred at the elevated critical height. (b) Profiles of the horizontal velocity perturbation for the inviscid solution of equation (1) with $\nu_e = 0$. Note that the perturbations become singular as the growth rate $c_i \to 0$; (i) In phase perturbations showing singular shearing over the crest and trough; (ii) Out of phase perturbations showing singular peak velocities over up/down slopes.

For a logarithmic velocity profile $\tan\theta = \varepsilon z_c/(z - z_c)$ and hence $\theta$ varies between

$$\theta \to 0 \quad \text{as } (z - z_c)/l_c \to \infty \quad \text{and} \quad \theta \to \pi \quad \text{as } (z - z_c)/l_c \to \infty \tag{5}$$

The imaginary part of the integral for $|z - z_c| \gg l_c$ then varies like

$$\text{Im}\{I\} \sim \varepsilon/c_i U'_c \theta = U''_c/U'^3_c \mathsf{H}(z - z_c) \quad \text{as } \xi \to \pm\infty, \tag{6}$$

where $\mathsf{H}(z - z_c)$ is the Heaviside step function. The result given by (6) is remarkable since it is independent of $c_i$ which means even for a slowly-growing wave it leads to an out of phase component of the motion that is independent of the growth rate, provided $\nu_e \neq 0$.

The significance of the term $iU''_c/U'^3_c$ in the solution for $I$ is that it yields an out of phase contribution to the vertical velocity that ultimately leads to the same contribution to the wave growth by the critical layer as found by Miles [1]. This result shows the solution found by Miles [1] is valid only when the waves grow sufficiently slowly such that

$$c_i \ll U'_c z_c \sim U_* \tag{7}$$

and hence the effects of the critical layer calculated by Miles [1] are valid *only* in the limit $c_i/U_* \downarrow 0$. We remark that Miles [1] only analysed the overall flow drag and energy input which required making a hypothesis about the singularity of the critical layer without considering the velocity profiles.

As fig. 2a shows, when $c_i \to 0$ the out of phase perturbation to $\Delta u$ becomes very large within a very thin layer of thickness of order $c_i/U'_c$. Effectively the vorticity in the $y$-direction $\omega_y$ is amplified on the lee side and reduced on the upwind side, which leads to the mean stream lines being deflected, with a lower pressure on the lee side and a higher drag.

Thus it is clear that in the limit of zero growth, i.e. $c_i \to 0$ (for small but finite $\nu_e$), the amplitude of the unsteady critical layer vorticity distortion mechanism tends to zero. However this was not the mathematical conclusion of Miles [1], who did not calculate the profile in the critical layer. He calculated (for a flow where $z_c > l_s$ i.e. $c_r > U_*$) the overall drag, $C_D$, and energy input, $E$, as a function of $c_i$ and the profile $U(z)$, and derived finite values for these overall properties of the flow for finite $c_i, \varepsilon$, and $a/L$. But he also deduced that $C_D$ and $E$ are finite as $c_i \to 0$. This is because he did not allow for the effects of small but finite eddy diffusivity, and consequently, as shown in the appendix, the contributions by the inertial critical layer to $C_D$ and $E$ are zero. Lighthill [16] provided an approximate physical analysis of the distortion of the vorticity produced by the wave (ignoring viscous effects and the inner surface layer) in the limit of $c_i \to 0$. The perturbed flow field with the jet on the lee slope shown in fig. 2b is consistent with his 'delta function' analysis (see p. 391 of reference [16]). But because he ignores the effect of finite eddy viscosity his physical conclusion is only correct when $c_i$ is finite.

In order to have a complete solution it is necessary to consider the surface shear layer and its asymptotic interaction with the whole flow.

## 3. Surface shear layer analysis for steady flows and matching with outer flow

The AML analysis of the surface layer and its matching with the outer region formally requires an asymptotic expansion for first and second order terms in the perturbation velocity, see fig. 3. For the longitudinal component, these are denoted by $\Delta u_0$, and $\Delta u_1$, i.e. $\Delta u = \Delta u_0 + \varepsilon \Delta u_1$ [22]. The perturbation equations for this layer are, like other boundary layer problems, the momentum equations with a perturbation pressure $\Delta p$, where $\Delta p$ denotes the ratio of the perturbation pressure to the density. For steady flow, in a frame of reference moving with the wave, for $z < l_s$

$$\mathscr{U}(z)\frac{\partial \Delta u_s}{\partial x} + \Delta w_s \frac{dU}{dz} = -\frac{\partial \Delta p}{\partial x} + \frac{\partial \Delta \tau}{\partial z} \qquad (8)$$

where $\Delta u_s, \Delta w_s$ and $\Delta p$ match with the outer region solution (discussed in section 2). To leading order the results are sensitive (to at least a factor of 2 in $\Delta u$) to the model for $\Delta \tau$ in relation to $\Delta u$. Within the surface layer, where the turbulence is in the local equilibrium turbulence, the usual mixing-length eddy viscosity model, leads to

$$\Delta \tau \sim z U_* \frac{\partial \Delta u}{\partial z}.$$

So that to leading order in (8) $\Delta \tau$ is negligible, but to first order $\Delta u_1(z)$ is determined by $\Delta \tau$. At the bottom of the layer, as with all non-uniform turbulent boundary layers, the rapid variation of $\partial \Delta\tau/\partial z$ (in fact logarithmic) determines the profiles of the perturbations, even if there is an 'immersed' critical layer [13]. Above the surface layer the turbulence is not in equilibrium, which leads to a lower value of $\Delta\tau$, and a higher value of speed-up $\Delta u$. Vertical profiles of the various terms in the turbulent energy equation for production, dissipation and transport over waves are reviewed by Belcher & Hunt [6]. Similarly, this application of AML can be extended to the intense local turbulence processes in the highly sheared, elevated critical layers over unsteady waves. This shows that the perturbation turbulent stresses in the outer region can be so significant that they can destroy the inertial effect of these critical layers.

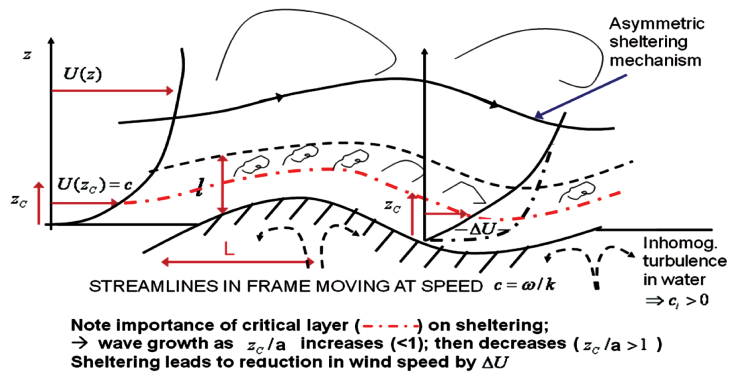

*Figure 3.* Schematic of wind over wave mechanisms for steady low amplitude waves showing sheltering mechanism in the surface layer and its coupling with the outer flow. The immersed critical layer has no significant effect.

For calculating such complex interacting flows, when the asymptotic layers are no longer distinct and interact significantly, it is more straightforward and more flexible to have a single differential equation for the perturbations, where the inhomogeneous, equilibrium and non-equilibrium turbulent stresses are modelled with suitable relaxation adjustments, see the appendix.

## 4. Approximate application of AML methods to complex wind wave models

Almost all photographs of ocean waves [14], and fig. 4 and even wind-driven capillary waves on ponds show that they form in groups of individual waves which vary in size from small to large and then small again, typically about 5 to 7 waves in the folk-lore of mariners. The fact that their form broadly persists over large distances compared with the size of the wave group is generally assumed to be a results of nonlinear interactions between waves in the water, reinforced by nonlinear interactions between the waves and the wind [23]. Persistent wave groups form both when the waves do not break and when they break systematically within the groups, in general on the down side of the group. A diagram of a typical wave group, showing the asymptotic layers in the wind flow is shown in fig. 5a. (see also [3]). In such a group the individual waves grow on the upwind side of the group and decrease in amplitude on the downwind side. Because of the asymmetry of the flow over the whole group, the critical layer $z_c$ is higher over the downwind than the upwind part of the wave group. As section 2 and fig. 2b demonstrate, this means that the net effect of the unsteady dynamics (i.e. positive drag from growing waves with lower $z_c$ exceeds the negative drag from the decreasing waves in the down-wind part) contributes to the mean drag and energy input to the wave group. In addition the higher value of $z_c$ thickens the surface layer and adds to the sheltering drag [13]. This hypothesis, based on applying AML concepts to this complex flow, needs to be tested over a wide parameter range of flow and wave groups.

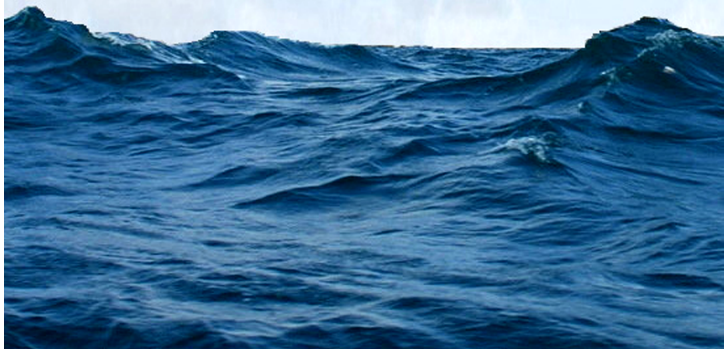

*Figure 4.* Wind over wave groups. Typical random groups of ocean waves (not breaking, but with sharp crests. Photograph taken off the coast of the eastern Mediterranean sea).

Initial computations of turbulent flows over specified groups of 3 dynamic waves, (which are rising and falling as they move), using a non-equilibrium eddy viscosity model [24][2] shows, from streamwise velocity profiles, for $U_\infty = 10.8$ m/s (fig. 5b), how $z_c$ is higher on the downwind than on the upwind part of the wave group. This, and also computations of streamline patterns, for $U_\infty = 2.18$ m/s (fig. 5c), are consistent with the hypothesis of asymmetric flow over the group. For all these simulations $|c_i|/c_r \doteq \frac{1}{10}$ being approximately equal to $a_{\max}/L_G$, where $a_{\max}$ denotes the maximum amplitude in the group and $L_G$ is the wavelength of the group.) As can be seen from these diagrams the critical layer height of $z_c$ varies over the wave groups. We emphasize the same anomaly is observed over unsteady (growing) monochromatic waves [24].

---

[2] The turbulence model adopted here is the high-Reynolds number extension of that given by Sajjadi et al. [31].

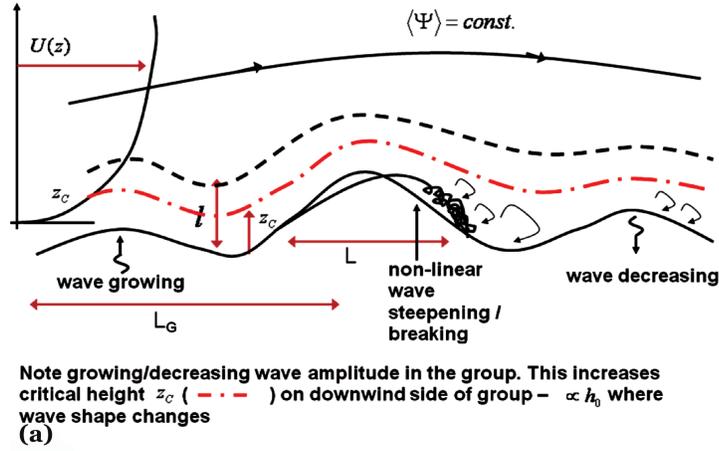
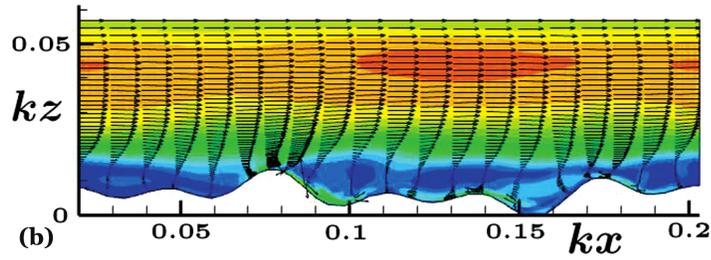
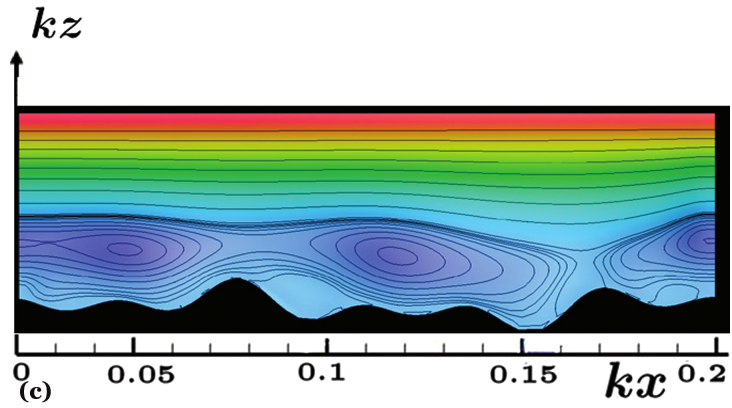

*Figure 5.* (a) Schematic of asymmetric wind flow over wave groups showing separation and changing behaviour of the critical layer. (b, c) Computations of turbulent flows over wave groups: (b) mean velocity, (c) mean streamlines relative to average velocity of the waves.

The consequence of this physical picture of turbulent wind-wave interactions is that the wave cannot be regarded as a random surface uncorrelated (on the scale of typical wave groups) with the wind structure. With remote sensing of waves and wind, and analysis of their correlations (e.g. with wavelet, as opposed to Fourier, methods), and AML methods applied to groups of waves, it should become possible to improve wave modelling and forecasting in future.

Another major challenge is to include the generalization effects of separation [25] and wave breaking [26]. This also provides a systematic approach for studying different types of wind-wave interactions for quite different forms of wind structure such as occurs in tropical cyclones [27,28], and when significant ocean currents affect the wave growth and wave groups [3,24].

## Appendix

### A. Effect of the inertial critical layer

In a frame of reference moving with the waves, the vertical perturbation to the air flow, $\Delta w = \mathscr{W}(z)e^{ik(x-c_r t)+kc_i t}$, satisfies the Orr-Sommerfeld-like equation [29,30][3]

$$\mathscr{T}'' \equiv (\nu_e \mathscr{W}'')'' = ik[(\mathscr{U} - ic_i)(\mathscr{W}'' - k^2 \mathscr{W}) - U''\mathscr{W}] \qquad (9)$$

where $\nu_e$ is the eddy viscosity.

In the outer region, turbulence is negligible and thus the left-hand side of (9) can be neglected compared to the right-hand side and thus we obtain the Rayleigh equation

$$(\mathscr{U} - ic_i)(\mathscr{W}'' - k^2 \mathscr{W}) - U''\mathscr{W} = 0 \qquad (10)$$

---

[3] Miles and Sajjadi arrived at the same equation independently, but they invoked different turbulence closure schemes for the turbulent flow above the surface waves.

As was shown by Sajjadi [32], the leading order solution to (10) is

$$\mathscr{W} = (\mathscr{U} - ic_i)e^{-kz}\left[A + \mathscr{W}_c U'_c e^{kz_c}\int_0^\infty\left\{\frac{1}{(\mathscr{U}-ic_i)^2}-1\right\}dz\right]\quad(11)$$

where $A$ is constant which can be determined by matching the solutions to the outer and the inner regions.

For slow growing waves $c_i > 0$, the critical layer lies within the inner region close to the surface wave and the integral in (11) is regular since $\mathscr{U} > 0$ there. Let us now suppose that

$$c_i \ll U'^2_c/U''_c$$

then the integral in (11) can be evaluated approximately.

Hence, indenting the path of integration in (11) under the singularity $z = z_c$, we obtain

$$\mathscr{W} = (\mathscr{U} - ic_i)e^{-kz}\left[A + \mathscr{W}_c U'_c e^{kz_c}\left(\oint_0^\infty\left\{\frac{1}{(\mathscr{U}-ic_i)^2}-1\right\}dz - I\right)\right]\quad(12)$$

where

$$I = \lim_{\varpi \to 0}\int_{\eta_c-\varpi}^{\eta_c+\varpi}\left\{\frac{1}{(\mathscr{U}-ic_i)^2}-1\right\}dz\quad(13)$$

Expanding $\mathscr{U}(z)$ as a Taylor expansion in the vicinity of the critical point, i.e.

$$\mathscr{U}(z) \sim \eta U'_c + \tfrac{1}{2}\eta^2 U''_c + O(\eta^3),\qquad \eta \equiv z - z_c,$$

setting $z = z_c \varpi e^{i\theta}$, where $\varpi \equiv c_i/U_* \ll 1$, and

$$\tan\theta = -c_i/U'_c\eta$$

then (13) becomes

$$\begin{aligned}I &\sim \frac{1}{U'^2_c}\left\{\lim_{\varpi\to 0}\int_{z_c-\varpi}^{z_c+\varpi}\frac{dz}{(z-z_c)^2}+i\pi\frac{U''_c}{U'_c}\right\}\\ &= \frac{i\pi U''_c}{U'^3_c}\end{aligned}\quad(14)$$

which is in agreement with the result obtained by Belcher et al. [21].

As also pointed out by Belcher et al. [21], for a logarithmic mean velocity profile $\tan\theta = \varpi z_c/(z-z_c)$. Hence $\theta$ varies between 0 and $\pi$ as $(z-z_c)/l_c$ tends to $\pm\infty$, respectively. Note that, the transition between these limiting values occurs across the layer of thickness $l_c = \varpi z_c$. Note also, the significance of the term $iU_c''/U_c'^3$ in the solution for $I$ is that it leads to an out of phase contribution to the wave induced vertical velocity. This gives rise to the same wave growth-rate as that of Miles [1] critical-layer model.

The result of the present analysis confirms the earlier finding [21] in that Miles [1] solution is *only* valid when the waves grow significantly slowly such that

$$c_i \ll U_c' z_c \sim U_* \tag{15}$$

Our analysis also shows that when inertial effects controls the behaviour around the critical layer, there is a smooth behaviour around the critical layer of thickness [21]

$$l_c \sim c_i/U_c' \sim z_c c_i/U_* \tag{16}$$

Hence this proves the effects of critical layer [1] are *only* valid in the limit $c_i/U_* \downarrow 0$.

To calculate the energy-transfer parameter due to critical layer, $\beta_c$, we let $\mathscr{W} = -\mathscr{V}\mathscr{M}$, where $\mathscr{V} = \mathscr{U} - ic_i$. Thus, (9) becomes

$$[\nu_e(\mathscr{V}\mathscr{M}'' + 2U'\mathscr{M}' + U''\mathscr{M})]'' = ik[(\mathscr{V}^2\mathscr{M}')' - k^2\mathscr{V}^2\mathscr{M}] \tag{17}$$

In the quasi-laminar limit the left-hand side of (17) is negligible and thus we have

$$(\mathscr{V}^2\mathscr{M}')' - k^2\mathscr{V}^2\mathscr{M} = 0 \tag{18}$$

Multiplying (18) by $\mathscr{M}$, integrating by parts over $0 < z < \infty$, and invoking the inner limits $\mathscr{M} \to a$ and $\mathscr{V}^2\mathscr{M}' \to \mathscr{P}_0$ (the complex amplitude of the surface pressure) and a null condition at $z = \infty$, we obtain

$$a\mathscr{P}_0 = -\int_0^\infty \mathscr{V}^2(\mathscr{M}'^2 + k^2\mathscr{M}^2)\,dz \tag{19}$$

Using the simplest admissible trial function for the variational integral (19), i.e.

$$\mathscr{M} = ae^{-kz/\varsigma} \tag{20}$$

where $\varsigma$ is a free parameter. Substituting (20) into (19) together with the approximation $\mathscr{V} \approx U_1 \ln(z/z_c) - ic_i$ we get

$$\hat{\mathscr{P}}_0 \equiv \mathscr{P}/kaU_1^2 = -k(\varsigma^{-2} + 1)\int_0^\infty e^{-2kz/\varsigma}\mathscr{F}(z)\,dz$$

where
$$\mathscr{F}(z) = \ln^2(z/z_c) - 2i\hat{c}_i \ln(z/z_c) - \hat{c}_i^2$$
and $\hat{c}_i = c_i/U_1$. Evaluating the integral we obtain

$$\hat{\mathscr{P}}_0 = -\frac{\varsigma + \varsigma^{-1}}{2}\left\{\frac{\pi^2}{6} + \ln^2\left(\frac{2\gamma\xi_c}{\varsigma}\right) - 2i\hat{c}_i \ln\left(\frac{2\gamma\xi_c}{\varsigma}\right) + \hat{c}_i^2\right\} \quad (21)$$

where $\xi_c \equiv kz_c$ (cf. [1], see also the caption of figure 1), $\gamma = 0.5772$ is Euler's constant, $U_1 = U_*/\kappa$, and $\kappa = 0.41$ is von Kármán's constant. It then follows from the variational condition $\partial\hat{\mathscr{P}}_0/\partial\varsigma = 0$ that

$$\varsigma^2 = \frac{L_\varsigma^2 - 2(1+i\hat{c}_i)L_\varsigma + (\hat{c}_i^2 + 2i\hat{c}_i + \pi^2/6)}{L_\varsigma^2 + 2(1-i\hat{c}_i)L_\varsigma + (\hat{c}_i^2 - 2i\hat{c}_i + \pi^2/6)} \quad (22)$$

where $L_\varsigma = -(L_0 + \ln\varsigma)$ and $L_0 = \gamma - \ln(2\xi_c) = \Lambda^{-1}$.

The corresponding critical-layer approximation to the energy-transfer parameter $\beta$ may then be calculated from (12), which implies $\mathscr{W}_c = \mathscr{P}_c/U_c' \approx \mathscr{P}_0/U_c'$, and (14), which yields

$$\begin{aligned}\beta_c &= \pi\xi_c|\mathscr{W}_c/U_1 a|^2 = \pi\xi_c^3|\hat{\mathscr{P}}_0|^2 \\ &= \tfrac{1}{4}\pi(\varsigma+\varsigma^{-1})^2\left|\left(L_\varsigma^2 - 2i\hat{c}_i L_\varsigma + \hat{c}_i^2 + \tfrac{1}{6}\pi^2\right)\right|^2 \\ &= \pi\xi_c^3 L_0^4\left[1 + \left(4 - \tfrac{1}{3}\pi^2 + 10\hat{c}_i^2\right)\Lambda^2 + \mathcal{O}(\Lambda^3)\right].\end{aligned} \quad (23)$$

To obtain the corresponding expression for the component of the energy-transfer parameter, $\beta_T$, due to turbulence, we multiply (9) by $-\mathscr{M}$, integrating over $0 < z < \infty$, invoking the conditions

$$\mathscr{M} = a, \quad \mathscr{M}' = ka, \quad \mathscr{T}' = ik[\mathscr{P}_0 - kac^2]$$

on $z = 0$ and the null condition for $z \to 0$, we obtain

$$\begin{aligned}\int_0^\infty \mathscr{M}\mathscr{T}''\,dz &= ka[\mathscr{T}_0 - i\mathscr{P}_0] + i(kac)^2 + \int_0^\infty \mathscr{M}''\mathscr{T}\,dz \\ &= i(kac)^2 + ik\int_0^\infty \mathscr{V}^2\left(\mathscr{M}'^2 + k^2\mathscr{M}^2\right)dz,\end{aligned} \quad (24)$$

where $\mathscr{T}_0$ is the complex amplitude of the surface shear stress and $c = c_r + ic_i$. Then, in the limit as $s \equiv \rho_a/\rho_w$, where $\rho_a$ and $\rho_w$ are densities of the air and water, respectively, we obtain from

$$\alpha + i\beta \equiv (c^2 - c_w^2)/sU_1^2 = (\mathscr{P}_0 + i\mathscr{T}_0)/kaU_1^2 \equiv (\hat{\mathscr{P}}_0 + i\hat{\mathscr{T}}_0), \quad (25)$$

where $c$ is the complex wave speed,

$$c_w = \sqrt{g/k} - 2ik\nu_w, \qquad |k\nu_w/c| \ll 1$$

is the speed of water waves in the absence of the airflow above it, $\nu_w$ is the kinematic viscosity of water, and the suffix zero denotes evaluation at $z = 0$. Then if follows from (24) and (21) that

$$\alpha_T + i\beta_T = (kaU_1)^{-2} \int_0^\infty \left\{ i\nu_e \left[ \mathscr{V} \mathscr{M}''^2 + 2U' \mathscr{M} \mathscr{M}'' + U'' \mathscr{M} \mathscr{M}'' \right] \right.$$
$$\left. - k\mathscr{V}^2 \left( \mathscr{M}'^2 + k^2 \mathscr{M}^2 \right) \right\} dz.$$

The above integral can be evaluated asymptotically[4] whose imaginary part yields

$$\beta_T = 5\kappa^2 L_0 + \mathcal{O}(\Lambda). \tag{26}$$

In fig. 6, we show comparison of the energy-transfer rate, $\beta$, between the present result for a monochromatic unsteady (growing) wave, both analytically and numerically, and those calculated by Miles [1] and Janssen [18] for the steady wave counterpart. Miles and Janssen both assume that the drag $C_D$, and thence $\beta$, is dominated by the limiting inviscid wave growth mechanism, thus their formulation is independent of $c_i$. In contrast, the present calculation is for a viscous unsteady (growing) wave, where $c_i/U_* = 0.01$ and $kz_0 = 10^{-4}$.

We emphasize that the various models [13,11,33] all generally agree with our numerical simulations performed using the Reynolds-stress closure scheme [34] for the energy transfer parameter, $\beta$, shown in fig. 6. This shows consistency between these models and the unimportance of very small $c_i$ for which viscous processes are significant.

We remark that, these parameterizations have been incorporated and tested in spectral wave models, WaveWatch and WindWave, which shows a superior results when compared with field data [35,36].

Figure 7 shows comparison of $\beta_c$ as a function of wave age $c_r/U_1$, calculated from the numerical solution of inviscid Orr-Sommerfeld equation [37], against the numerical solution of equation (1) for $c_i/U_* = 0.01$, $kz_0 = 10^{-4}$ and $\nu_e \neq 0$. We remark that increasing $c_i/U_*$ from 0.01 to 0.1 (not shown here) makes no significant difference in the magnitude of $\beta_c$. We conclude therefore for a finite value of $\nu_e$ the right-hand side of equation (1) is dominant and therefore the magnitude of $\beta_c$, calculated from the solution of (1), is practically zero over a wide range of the wave age, in particular for a 'young' wave, where $c_r/U_1 < 2$. We thus conclude that the critical-layer mechanism plays an insignificant role for $c_r/U_1 < 9$, and very little effect for $9 \leq c_r/U_1 \leq 10.5$.

---

[4] The detailed evaluations may be obtained from the authors upon request.

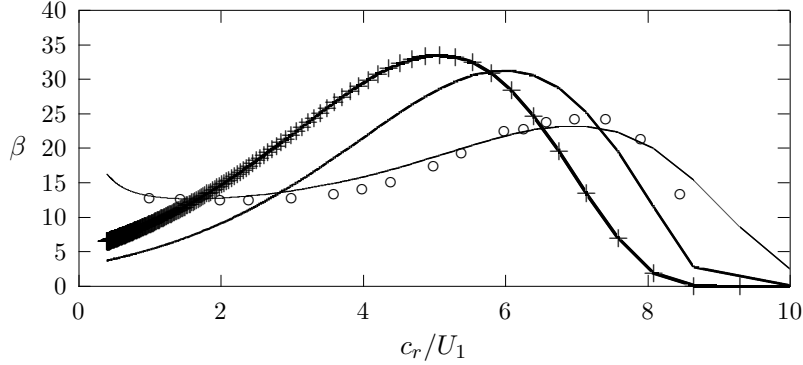

*Figure 6.* Total energy transfer parameter, $\beta$, due to the combined effect of sheltering and inertial critical layer for growing waves (where $c_i \ll U_*$) as a function of the wave age $c_r/U_1$. +++++, Miles [1] calculation ($c_i = 0, \nu_e = 0$) from his formula: $\beta = \pi \xi_c \left\{ \frac{1}{6}\pi^2 + \log^2(\gamma \xi_c) + 2\sum_{n=1}^{\infty} \frac{(-1)^n \xi_c^n}{n! n^2} \right\}^2$, where $\xi_c = kz_c$ is the critical height $\xi_c = \Omega(U_1/c_r)^2 e^{c_r/U_1}$ and $\Omega = gz_0/U_1^2$ is the Charnock's constant [38].
Thick solid line, parameterization of Miles formula [18], for $c_i = 0, \nu_e = 0$: $\beta = 1.2\kappa^{-2}\xi_c \log^4 \xi_c$, where $\xi_c = \min\left\{1, kz_0 e^{[\kappa/(U_*/c+0.011)]}\right\}$.
Thin solid line, present formulation: $(\beta_T + \beta_c)$ for $c_i \neq 0, \nu_e \neq 0$. ∘, Numerical simulation using the Reynolds-stress closure model [34] for $c_i \neq 0, \nu_e \neq 0$. Note that, $\beta$ given in [1,18] is equivalent to $\beta_c$ in our notation.

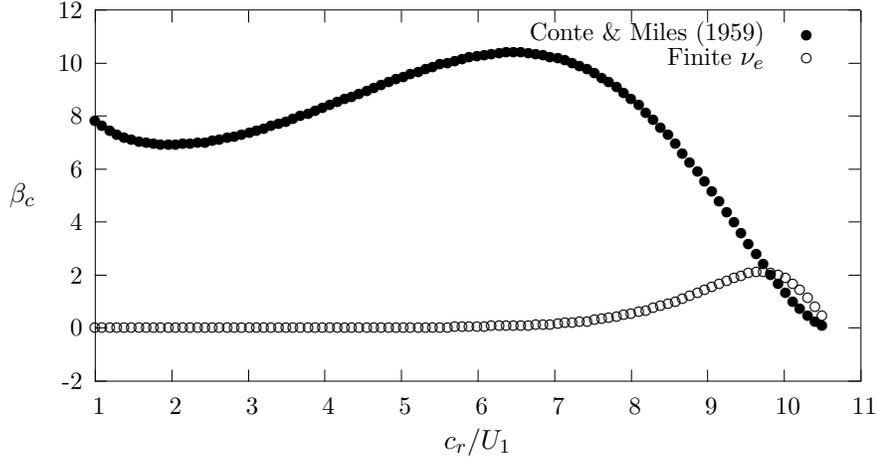

*Figure 7.* Component of energy transfer parameter, $\beta_c$, due to inertial critical layer for growing waves (where $c_i \ll U_*$) as a function of the wave age $c_r/U_1$. •, numerical solution of inviscid Orr-Sommerfeld equation [37] for $c_i = 0$ and $\nu_e = 0$ using the singular critical layer approach; ∘ numerical solution of equation (1) for $c_i \neq 0$ and $\nu_e \neq 0$.

**Acknowledgements**

We would like to thank the referees for the their critical review and useful comments which has substantially improved the manuscript.